\documentclass[pra, amsmath, reprint, amssymb, aps, superscriptaddress, twocolumn]{revtex4-1}
\usepackage{graphicx}
\usepackage{amsmath, braket, amsfonts}

\usepackage{multirow}
\usepackage{mathrsfs,amsmath} 
\usepackage{fixmath}
\usepackage{graphicx}
\usepackage[colorlinks,linkcolor=blue,anchorcolor=red,citecolor=green]{hyperref}
\usepackage{placeins}
\usepackage{cancel}

\renewcommand\vec{\boldsymbol}

\begin{document}


\title{Probing the Quantum Noise of the Spinon Fermi Surface with NV Centers}

\author{Jun Yong Khoo}
\affiliation{Institute of High Performance Computing, Agency for Science, Technology, and Research, Singapore 138632}
\affiliation{Max-Planck-Institut für Physik komplexer Systeme, Nöthnitzer Straße 38, 01187 Dresden, Germany}

\author{Falko Pientka}
\affiliation{
Institut f\"ur Theoretische Physik, Goethe-Universit\"at, 60438 Frankfurt a.M., Germany
}
\affiliation{Max-Planck-Institut für Physik komplexer Systeme, Nöthnitzer Straße 38, 01187 Dresden, Germany}

\author{Patrick A. Lee}
\affiliation{Department of Physics, Massachusetts Institute of Technology, Cambridge Massachusetts 02139, USA}

\author{Inti Sodemann Villadiego}
\affiliation{Institut für Theoretische Physik, Universität Leipzig, Brüderstraße 16, 04103, Leipzig, Germany}
\affiliation{Max-Planck-Institut für Physik komplexer Systeme, Nöthnitzer Straße 38, 01187 Dresden, Germany}
\date{\today}

\begin{abstract}
    We study the transverse electrical conductivity and the corresponding magnetic noise of a two-dimensional U(1) spin liquid state with a spinon Fermi surface. We show that in the quasi-static regime these responses have the same wave-vector dependence as that of a metal but are reduced by a dimensionless pre-factor controlled by the ratio of orbital diamagnetic susceptibilities of the spinons and chargons, correcting previous work. We estimate that this quasi-static regime is comfortably accessed by the typical NV center splittings of a few GHz and estimate that the expected $T_1$ times for an NV center placed above candidate materials, such as the organic  dmit and ET salts,  monolayer 1T-TaS$_2$/Se$_2$, would range from several tens to a few hundred milliseconds.
\end{abstract}



\maketitle

\section{Introduction}

A recent work~\cite{Khoo2021} has shown that even when a metallic Fermi surface has a complex shape, the dissipative real part of the transverse conductivity is a universal quantity controlled entirely by its geometric shape, which in turn gives rise to a universal $T_1$ time for an NV center controlled only by the length of the perimeter of the Fermi surface (in the regime where the NV center distance $z$ satisfies $p_{\rm F}^{-1} \ll z \ll l_{\rm mfp}$). However, while the conclusions of Ref.~\cite{Khoo2021} on metallic Fermi surface states remain valid, the same work concluded incorrectly that the spinon Fermi surface state would have the same value for the low-frequency limit of the transverse conductivity and the associated $T_1$ time as that of a metallic Fermi liquid. 

The reason for this discrepancy is that
Ref.~\cite{Khoo2021} inadvertently missed the diamagnetic contributions of the spinons. As we will discuss here, the spinon transverse conductivity assumes a standard form of a Fermi liquid given by (in the limit $\omega\ll v_{\rm F} q \ll \epsilon_{\rm F}$)\cite{Pines,GiuVig}


\begin{eqnarray}\label{sigmaFL}
\sigma^{\rm spinon}_{\perp} (q,\omega) \approx \frac{e^2}{h} \frac{p_{\rm F}}{q}+ \chi_s \frac{q^2}{i \omega}.
\end{eqnarray}

%
Moreover, we will show that the expression for the physical transverse {\it electrical} conductivity of the spinon Fermi surface state (in the low frequency limit) is given by:
\begin{eqnarray}\label{sigmaSF}
\sigma _{\perp} (q,\omega) \approx \left( \frac{\chi_c}{\chi_s+\chi_{a}+\chi_c} \right)^2 \frac{e^2}{h} \frac{p_{\rm F}}{q}+ \chi \frac{q^2}{i \omega}.
\end{eqnarray}
The above expressions apply in the zero temperature and clean limit of a spin-degenerate, circular Fermi surface with radius $p_{\rm F}$. $\chi_{s,c}$ are spinon and chargon orbital diagmagnetic susceptibilities, $\chi_{a}$ is a diamagnetic susceptibility associated with the Maxwell term of the emergent U(1) gauge field and $\chi^{-1}=(\chi_s+\chi_{a})^{-1}+\chi^{-1}_c$. Therefore, we see that while the transverse conductivity has the same form as that of an ordinary metal in this quasi-static regime, its real (dissipative) part is corrected by the dimensionless factor $\chi_c^2/(\chi_s+\chi_{a}+\chi_c)^2$. As a consequence, the remarkable result of Ref.~\cite{Khoo2021} that the relaxation rate $1/T_1$ of an NV center coupled to a two-dimensional spinon Fermi surface state has the same dependence on distance, frequency, and temperature as in the metallic case remains valid, albeit with an extra pre-factor, $\chi_c^2/(\chi_s+\chi_{a}+\chi_c)^2$ (see Table~\ref{Table.spinon} for specific estimates).

\section{Spinon Fermi surface low-energy theory}

The following low energy effective Lagrangian captures the orbital coupling of the spinon Fermi surface state to the physical electromagnetic fields (denoted by $\vec{E} = -\vec{\partial}_{\vec{r}} \varphi - \partial_t \vec{A}$ and $\vec{B} = \vec{\partial}_{\vec{r}} \times \vec{A}$):

\begin{multline}\label{Lag}
\mathcal{L} = \mathcal{L}_{\rm spinon} (\vec{p}-\vec{a})+\mathcal{L} _{\rm chargon} (\vec{p}-\vec{A}+\vec{a})\\ +\frac{\epsilon_{a}}{2}\vec{\mathsf{e}}^2  - \frac{\chi_{a}}{2} \vec{\mathsf{b}}^2 + \cdots .
\end{multline}

%
%
Let us briefly explain the physical meaning of the various terms. Here we are imagining that deconfinement of the U(1) gauge field has taken place and thus the emergent  electric and magnetic fields, denoted by $\vec{\mathsf{e}} = -\vec{\partial}_{\vec{r}} \phi - \partial _t \vec{\mathsf{a}}$ and $\vec{\mathsf{b}} = \vec{\partial}_{\vec{r}} \times \vec{\mathsf{a}}$, can be taken to be non-compact. The emergent gauge field now acquires dynamics which is described by a Maxwell-like term and parametrized by emergent dielectric and diamagnetic constants $\epsilon_{a}$ and $\chi_{a}$. 
$\mathcal{L} _{\rm spinon}  (\vec{p}-\vec{a})$ is a short-hand for the Lagrangian of a fermion (the spinon) minimally coupled to the emergent U(1) gauge field $\vec{a}$, but neutral under the physical gauge field $\vec{A}$. On the other hand $\mathcal{L} _{\rm chargon} (\vec{p}-\vec{A}+\vec{a})$ describes the Lagrangian of a boson (the chargon) that minimally couples to the physical field and carries opposite gauge charge to the spinon with respect to the emergent field. 
The above Lagrangian can be motivated from a slave-boson (or the closely related slave-rotor~\cite{florens2004slave}) parton decomposition of the electron operator which is viewed as the composite of the spinon and chargon (see e.g.~\cite{lee2006doping,senthil2008theory}). Nevertheless, we would like to emphasize that in the above Lagrangian the spinon and chargon should not be viewed strictly speaking as the same objects as the unphysical UV partons, but rather as low energy deconfined physical quasiparticles. In fact, due to the presence of the Maxwell term, if we were to change the Lagrangian in Eq.(3) by coupling the spinon to the physical gauge field instead of the chargon, one would obtain different predictions for phyisical gauge-invariant observables. Therefore, in this sense, we no longer have the freedom to assign the coupling of the electromagnetic field to either the chargon or the spinon and get the same answer.

Because the chargons are gapped they can be integrated out for the purpose of the low energy description, and thus their lagrangian can be replaced by a Maxwell term of the form:

\begin{align}\label{Lchar}
\mathcal{L} _{\rm chargon}  (\vec{p}-\vec{A}+\vec{a}) \approx   \frac{\epsilon _c}{2} (\vec{\mathsf{e}}-\vec{E})^2  - \frac{\chi_c}{2} (\vec{\mathsf{b}}-\vec{B})^2.
\end{align}

The effective dielectric and diamagnetic constants of the chargons, $\epsilon_c$ and $\chi_c$, will depend on their detailed microscopic dispersion. For example, within a relativistic boson model of the chargon dispersion, one obtains $\chi_c=1/(24\pi m_c)$, where $m_c= \Delta_c/v_c^2$, is the effective mass of the chargons with a speed $v_c$ and a gap $\Delta_c$, as shown in Ref.~\cite{dai2020modeling}.

On the other hand for the spinons, one can keep track of their occupation of momentum states only within a narrow sliver around the Fermi surface while integrating out higher energy modes. This  amounts to replacing $\mathcal{L} _{\rm spinon}$ by the following Lagrangian:
\begin{align}\label{Leff}
\mathcal{L} _{\rm spinon}  (\vec{p}-\vec{a}) \approx \mathcal{L}^{FS} _{\rm spinon} (\vec{p}-\vec{a}) + \frac{\epsilon _s}{2} \vec{\mathsf{e}}^2  - \frac{\chi_s}{2} \vec{\mathsf{b}}^2.
\end{align}
The Maxwell terms for the emergent electromagnetic fields arise from integrating out the modes away from the Fermi surface, and in particular the magnetic term accounts for the spinon diamagnetism ($\chi_s$). These terms were missed in Ref.~\cite{Khoo2021}. For a non-relativistic parabolic dispersion for the spinons one obtains $\chi_s=g_s/(24\pi m_s)$ (namely the standard Landau diagmagnetic constant with $g_s=2$ accounting for the spin degeneracy).

Next we follow the classic work of Ioffe and Larkin~\cite{Ioffe1989} to derive a relation between the physical conductivity and the spinon and chargon conductivities. Due to the appearance of the Maxwell term in Eq.~\eqref{Lag}, which was absent in the original Ioffe-Larkin paper, the derivation is a little different and the result slightly modified. First, the Euler-Lagrange equations of motion that follow from Eq.~\eqref{Lag}, by taking its variational derivatives with respect to $\vec{a}$, $\delta \mathcal{L}/\delta \vec{a}=0$, is given by:

\begin{equation}\label{dldA}
\chi_{a} \vec{\partial}_{\vec{r}} \times  \vec{\mathsf{b}} - \epsilon_{a} \partial _t   \vec{\mathsf{e}} =
 \vec{j}_{\rm s} - \vec{j}_{\rm c}.
\end{equation}

\noindent where ${\vec j}_{s}=\delta\mathcal{L}_{\rm spinon}/\delta \vec{a}$ and ${\vec j}_c=-\delta\mathcal{L}_{\rm chargon}/\delta \vec{a}$ are the spinon and chargon current densities respectively. Second, by taking functional derivatives with respect to the physical gauge field, we find that:

\begin{equation}\label{ecurrent}
{\vec j}_e=\frac{\delta\mathcal{L}}{\delta\vec{A}}={\vec j}_c,
\end{equation}

\noindent so that the physical electric current density ${\vec j}_e$ is the same as the chargon current ${\vec j}_c$ but is modified from the spinon current. 
To obtain explicit dependence in the electric fields, 
we use Faraday's law $\vec{\partial}_{\vec{r}} \times \vec{\mathsf{e}}= -\partial _t  \vec{\mathsf{b}}$ to eliminate $\vec{b}$ in Eq.~\eqref{dldA}. It is convenient to decompose the currents into perpendicular and transverse components and define the corresponding conductivity as:

\begin{align}
\begin{split}\label{perpconducs}
&{\vec j}_{e\perp}=\sigma_\perp \vec{E}_{\perp},\\
&{\vec j}_{s\perp}=\sigma_{s\perp} \vec{\mathsf{e}}_{\perp},\\ 
&{\vec j}_{c\perp}=\sigma_{c\perp} (\vec{E_{\perp}-\mathsf{e}}_{\perp}),\\ 
\end{split}
\end{align}

\noindent and similarly for the longitudinal components. Going to Fourier space, we find the relations between $\vec{e}$ and $ \vec{E}$:

\begin{align}
\begin{split}\label{evsE}
&(\sigma_{c\|}+\sigma_{s\|}+i\epsilon_{a} \omega)\vec{\mathsf{e}}_{\|}=\sigma_{c\|} \vec{E}_{\|},\\
&\left(\sigma_{c\perp}+\sigma_{s\perp}+i\epsilon_{a} \omega+\chi_{a} \frac{q^2}{i \omega}\right)\vec{\mathsf{e}}_{\perp}=\sigma_{c\perp} \vec{E}_{\perp},\\ 
\end{split}
\end{align}

\noindent here the conductivities are understood to be frequency and wave-vector dependent. 
It is convenient to introduce: 

\begin{align}
\begin{split}\label{}
&\sigma'_{s\|}\equiv \sigma_{s\|}+i\epsilon_{a} \omega,\\
&\sigma'_{s\perp}\equiv \sigma_{s\perp}+i\epsilon_{a} \omega+\chi_{a} \frac{q^2}{i \omega}.
\end{split}
\end{align}
%
Then, by using Eqs.~\eqref{ecurrent} and~\eqref{perpconducs} to compute $\sigma$ and using Eq.~\eqref{evsE} to eliminate $ \vec{\mathsf{e}}$, we arrive at a result with the same form as the classic Ioffe-Larkin formula, $\sigma^{-1}=\sigma_c ^{-1}+\sigma_s'^{-1}$. The  longitudinal and transverse components of the physical conductivity $\sigma$ are displayed below:
\begin{align}
&\sigma  _\| ^{-1}(q,\omega) = \sigma _{c\|}^{-1} (\omega) + (\sigma _{s\|}(q,\omega)+i\omega \epsilon_{a} )^{-1},\label{Eq.condsppll}\\
&\sigma  _\perp ^{-1}(q,\omega) = \sigma _{c\perp}^{-1} (q, \omega) + \left(\sigma _{s\perp}(q,\omega)+i\omega \epsilon_{a}+\chi_{a} \frac{q^2}{i \omega}\right)^{-1}\label{Eq.condspperp}
\end{align}
Compared with the stand Ioffe-Larkin formula,  the spinon conductivity has  extra terms which are proportional to $\epsilon_{a}$ and $\chi_{a}$.

Now we apply Eq.~\eqref{Eq.condspperp} to the problem at hand.  Since we are interested in the limit $\omega\ll v_{\rm F} q \ll \epsilon_{\rm F}$)  we can drop the term $i\omega \epsilon_{a} $ 
in Eq.~\eqref{Eq.condspperp}. We first treat the spinon as a free Fermi liquid and  use the form given by Eq.~\eqref{sigmaFL} for $\sigma _{s\perp}$. Since the chargon is gapped, we set $\sigma_{c\perp}=\chi_c \frac{q^2}{i \omega}$. Plugging  into Eq.~\eqref{Eq.condspperp} we obtain our main result given by Eq.~\eqref{sigmaSF}. The first term is the real (dissipative) part that enters into the calculation of the relaxation rate of the NV center. As noted in the introduction, it is reduced from that of a metal by the factor $\chi_c^2/(\chi_s+\chi_{a}+\chi_c)^2$. We can interpret this result as originating from the fact that the physical fluctuating current is reduced from that of a free Fermi sea by the factor $F=\chi_c/(\chi_s+\chi_{a}+\chi_c)$ because its flow is restricted by coupling to the emergent gauge field. The conductivity is suppressed by $F^2$ because it is proportional to  a product of two current operators. It is also worth noting that when compared with the standard Ioffe-Larkin formula, $\chi_s$ is replaced by $\chi_s+\chi_{a}$. Thus the appearance of the Maxwell term in the gauge field Lagrangian can be absorbed as a re-definition of the spinon diamagnetic susceptibility if one chooses to use the original Ioffe-Larkin formula.

In Ref.~\cite{Khoo2021} the spinon conductivity was treated more accurately including Fermi liquid corrections due to the scattering of quasi-particles near the Fermi surface as well as accounting for Landau parameters. For completeness we give the relevant formulas below which can be plugged into Eq. \eqref{Eq.condspperp} to give the physical conductivity. Nevertheless, as we will see, in the low-frequency limit  $\omega\ll v_{\rm F} q \ll \epsilon_{\rm F}$ and clean limit $q \gg l_{\rm mfp}^{-1}$ (see upcoming discussion in Section~\ref{Sec.spinonFStranscond}), the same reduction factor $F^2$ appears in the dissipative part of the conductivity which enters into the relaxation rate of the NV center, and therefore one recovers the same limit as in Eq.~\eqref{sigmaSF} even within this more accurate description of the spinon conductivity.

%

%


%

The contribution to the spinon density ($\rho _{\rm FS}$) and current ($\vec{j}_{\rm FS}$) arising from fluctuations in the vicinity of the Fermi surface can be expressed in terms of the change of the spinon occupation near the Fermi surface ($\delta n_{\vec{p}}$) as follows:

\begin{eqnarray}
\rho _{\rm FS} &=& \frac{1}{\mathcal{A}} \sum _{\vec{p}}\delta n_{\vec{p}}, \ \vec{j}_{\rm FS} = \frac{1}{\mathcal{A}} \sum _{\vec{p}}\vec{v}_{\vec{p}} \delta \bar{n}_{\vec{p}}, \\
\delta \bar{n}_{\vec{p}} &=& \delta n_{\vec{p}} + \sum _{\vec{p'}} f_{\vec{p}\vec{p'}}\delta (\epsilon _{\vec{p}} - \epsilon_{\rm F}) \delta n_{\vec{p'}},
\end{eqnarray}
where $\vec{v}_{\vec{p}}=\partial_{\vec{p}} \epsilon({\vec{p}})$ is the spinon quasiparticle velocity and $f_{{\vec{p}},{\vec{p'}}}$ accounts for Landau parameters. The spinon distribution function obeys the linearized kinetic equation
\begin{align}
\partial _t \delta n_{\vec{p}} &+ \vec{v}_{\vec{p}} \cdot \vec{\partial}_{\vec{r}}\delta \bar{n}_{\vec{p}} +   \vec{\mathsf{e}} \cdot \vec{v}_{\vec{p}} \delta (\epsilon _{\vec{p}} - \epsilon_{\rm F})
= I [\delta n_{\vec{p}}],\label{Eq.spinoneom2}
\end{align}
where $I [\delta n_{\vec{p}}]$ accounts for momentum relaxing and momentum preserving collisions with respective scattering rates denoted by $\Gamma_1$ and $\Gamma_2$ based on the model for collisions described in Refs.~\cite{Khoo2021,LevGregNat,LevGregPNAS,Alekseev2018}. By solving the kinetic equation for a circular Fermi surface, one obtains the following expressions for the different conductivities:





\begin{align}
&\sigma _{s\|} = i\frac{n }{m}\left[\frac{2}{\frac{2in}{m} \rho _*(q,\omega) + F_1 \omega _- - \omega _+ - 2 i \Gamma _2 }\right]+i\omega \epsilon_s,\\
&\sigma _{s\perp}  = i\frac{n }{m} \left[\frac{2}{ F_1 \omega _-  - \omega _+  - 2i \Gamma _2 }\right]+i\omega \epsilon_s+\chi_s \frac{q^2}{i \omega}, \\
&\rho_* (q, \omega) =
- i\frac{1}{ n^2 \kappa} \frac{q^2}{\omega }, \quad \kappa = \frac{1}{n E_{\rm F}} \frac{1}{1 + F_0},\\
&\omega _\pm = \omega-i (\Gamma_1 +\Gamma_2) \pm \sqrt{\left[\omega-i (\Gamma_1 +\Gamma_2)\right] ^2 -(v_{\rm F}q)^2}, \\
&\sigma _{c\|}  = i\omega \epsilon_c, \ 
\sigma _{c\perp} = i\omega \epsilon_c+\chi_c \frac{q^2}{i \omega}, \label{Eq.condchperp}
\end{align}

%
\noindent where $n = p_{\rm F}^2/ 4\pi $ is the spinon density, $E_{\rm F} = p_{F}^2 /2m^*$ is the Fermi energy, and we have included only two non-zero Landau parameters $F_{0}$ and $F_1$. The Landau parameter $F_1$ determines the renormalization of the spinon quasiparticle mass $m^*$ relative to its transport mass $m$: $m = m^*/(1+F_1)$. The above expressions from Eqs.~\eqref{Eq.condchperp} can then be plugged onto Eqs.~\eqref{Eq.condspperp}. 

Although we are focusing on the properties of the transverse conductivity in this study, we mention in passing that the optical conductivity peaks at some typical scale that can be viewed as the optical pseudo-gap associated with the Mott scale (see Fig.~\ref{Fig.condspinon} and further information in Appendix~\ref{Sec.Mottgap}). The expression for this scale is:

\begin{equation}
\bar{\omega} = \sqrt{\frac{n }{m (\epsilon_c+\epsilon_s+\epsilon_{a})}}. 
\end{equation}

\section{Conductivities and magnetic noise of the Spinon Fermi surface state}~\label{Sec.spinonFStranscond}

\begin{figure}
\includegraphics[scale=1.0]{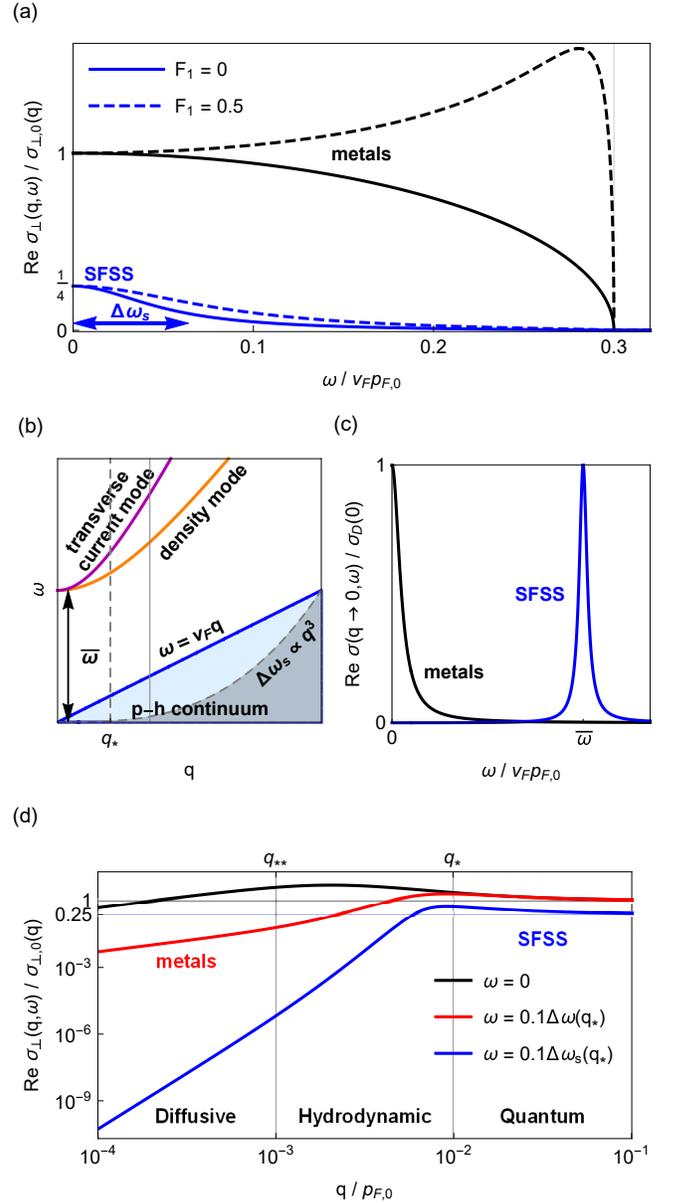}
\caption{
Comparison of the conductivity of the spinon Fermi surface state (labeled SFSS in blue) and a metal (black). This corrects Fig.~3 of Ref.~\onlinecite{Khoo2021} by including the spinon Landau diamagnetism. (a) Re~$\sigma  _\perp (q, \omega)$ for $q = 0.3 p_{\rm F,0}$ [along solid vertical line cut in (b)].
(b) Dispersion of collective modes and particle-hole excitations in the spinon Fermi surface state ($F_l = 0$). 
(c) Re~$\sigma _\| (q \rightarrow 0, \omega)$ with weak collisions: $\Gamma _1 = \Gamma _2 = 0.1 v_{\rm F} p_{\rm F,0}$. 
(b,c) The optical pseudo-gap in this case is $\bar{\omega} $ [see Appendix.~\ref{Sec.Mottgap} of the Supplementary Material~\cite{supplementary} for details]. 
(d) Re~$\sigma _\perp (q, \omega)$ at different (small) frequencies. We take here $\chi_s+\chi_{a}=\chi_c$ and $\epsilon_s=\epsilon_{a}=0$.
}
\label{Fig.condspinon}
\end{figure}

As discussed in Ref.~\cite{Khoo2021}, the clean limit of the quasi-static conductivity is attained for wave-vectors above an inverse spinon mean-free path, $q \gg l_{\rm mfp}^{-1}$, given by
\begin{align}
l_{\rm mfp}^{-1} &= {\rm max} (q_C, q_D),\\
v_{\rm F} q_D &=\frac{2\Gamma _1}{(1+F_1)},\ v_{\rm F}q_C = (\Gamma _1 + \Gamma _2).
\end{align}
In this wave-vector regime, Eq.~(\ref{sigmaSF}) can be obtained by taking the limit $\omega \rightarrow 0$, from the expressions in Eqs.~\eqref{Eq.condsppll}-\eqref{Eq.condchperp}. At fixed wave-vector but finite frequency, the real part of the transverse conductivity vanishes above a typical frequency window $\Delta \omega  _s $, as illustrated in Fig.~\ref{Fig.condspinon}, which can be estimated to be:
\begin{eqnarray}\label{Eq.spinonpeakwidth}
\frac{\Delta \omega _s}{2\pi}
&\simeq&\frac{(\chi_c+\chi_s+\chi_{a})q^3}{p_{\rm F}}.
\end{eqnarray}
Hence, measuring the quasi-static conductivity requires experiments to operate at frequencies $\omega \ll  \Delta \omega_s$. 
NV centers have a level splitting of about $3$GHz~\cite{Casola2018}, well below typical values for $\Delta \omega_s$ listed in Table~\ref{Table.spinon}, making them ideally suited to probe the quasi-static regime of the conductivity. As we will see later on, the main challenge is their rather weak coupling to the spin liquid which leads to relatively long $T_1$ times.

\begin{table}[t]
\begin{tabular}{lccccc}
\hline \hline
 & $k_{\rm F}^{-1}$ 
  & $\epsilon_{\rm F}$
& $\omega _p$
& $\Delta \omega _s/ 2\pi$ & $T_1$ \\ [-.2em]
& (A) & (meV) & (meV) & ${\rm (GHz)}$ & $({\rm ms})$ \\ \hline 
dmit & 2.4 & 59 & 80~\cite{Rao2019} & 215 & 58 \\ 
$\kappa$-ET & 3.2 & 98 & 87~\cite{Rao2019} & $2 \times  10^3$ & 78 \\ 
1T-TaS$_2$/Se$_2$ & 4.4 & - & 200~\cite{Rao2019} & -  & 109 \\ 
WTe$_2$ & 26 & 29 & 60~\cite{Wang2021} & $6 \times  10^4$  & 630 \\ 
\hline \hline
\end{tabular}
\caption{Order-of-magnitude estimates for material candidates (dmit stands for EtMe$_3$Sb[Pd(dmit)$_2$]$_2$ and $\kappa$-ET for $\kappa$-(ET)$_2$Cu$_2$(CN)$_3$). We took  $c = 0.5v_{\rm F}$, $F_1=0$ and $\chi_c=\chi_s+\chi_{a}$, and for simplicity $\epsilon _s=\epsilon _a = 0$, so that the Mott scale, $\bar{\omega}$, is determined by the chargon dielectric constant $\epsilon_c$. We have taken the temperature to be $T=10K$, and an NV center-to-sample distance of $z=1nm$ and the NV center splitting to be $\omega=3$GHz (we take $q=1/z$ to estimate $\Delta \omega _s$). The $\epsilon_{\rm F}$ and $\Delta \omega _s$ of 1T-TaS$_2$ are left blank as they are still uncertain (see discussion in main text). These values correct those of Table.~2 of Ref.~\onlinecite{Khoo2021}. 
Notice that the above are estimates for a single layer of each material. For multi-layered or bulk samples we can add up the contributions of each layer and there will be reduction of the $T_1$ time by a factor of the order $T_1 \rightarrow   T_1\frac{d}{z \log(l_{\rm mfp}/z)} \sim  T_1\frac{d}{z}$, where $d$ is the interlayer distance (assuming $l_{\rm mfp}\gg z\gg d$).}\label{Table.spinon}
\end{table}

In the more general case of a noncircular spinon Fermi surface, the real part of the quasi-static transverse conductivity (at low temperatures and for $p_{\rm F}^{-1}\gg q \gg 1/l_{\rm mfp}$) is given by:

\begin{equation}
{\rm Re}~\sigma _{\perp} = \left( \frac{\chi_c}{\chi_s+\chi_{a}+\chi_c} \right)^2 (2S+1)
\frac{e^2}{2 h q}\sum_i \mathcal{R}_{\rm F} |_{\vec{p}_i^* (\hat{q})}, \label{Eq.anisigperp0scatt}
\end{equation}
where $(2S+1)$ is the spin degeneracy factor, $\left\lbrace\vec{p}^*_i\right\rbrace$ is the set of points on the Fermi surface at which the Fermi velocity is orthogonal to $\hat{q}$, and $\mathcal{R}_{\rm F} |_{\vec{p}_i^* (\hat{q})} $ the absolute value of the local radius of curvature of the Fermi surface at $\vec{p}_i^*$.
Therefore, the conductivity only depends on the geometry of the Fermi surface and is reduced by the same factor, $\chi_c^2/(\chi_s+\chi_{a}+\chi_c)^2$, relative to the case of metals described in Ref~\cite{Khoo2021}. 

The above quasi-static transverse conductivity can be probed by measuring the $T_1$-time of a single spin (NV center) placed above the sample at distance $z$~\cite{Casola2018,Agarwal,Chatterjee2019}. This $T_1$-time is inversely proportional to the imaginary part of the magnetic field autocorrelation function (magnetic noise) at the NV center location, ${\rm Im}\chi _{B_\mu B_\nu} (z, \omega)$, and at the frequency $\omega$ given by the energy splitting of the NV center~\cite{Casola2018,Langsjoen2012}. For the circular Fermi surface, we have:
\begin{align}
{\rm Im} \chi _{B_z B_z}  (z, \omega)
&= \frac{\mu _0 ^2 \omega}{8 
\pi}\int q d q e^{-2qz} {\rm Re}~\sigma _{\perp} (q,\omega).
\end{align}

\begin{figure}[t]
\includegraphics[scale=1.0]{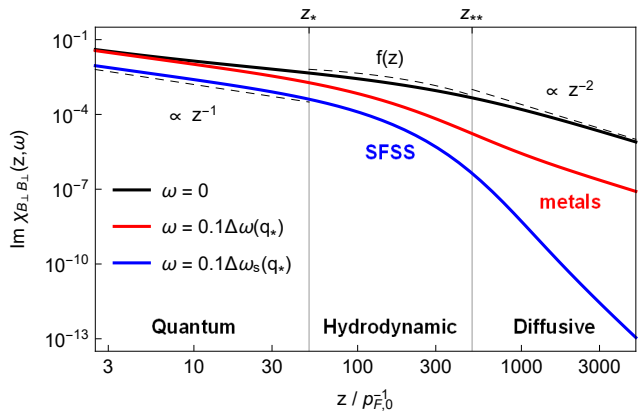}
\caption{
Magnetic noise of the spinon Fermi surface state (in blue and labeled SFSS) analogous to those shown in Fig.~5 of Ref.~\onlinecite{Khoo2021}, but including the Landau diamagnetism, for  $\epsilon_s=\epsilon_{a}=0$ and $\chi_s=\chi_c$. The analogous behavior of the metals is shown in the black and red curves. 
}
\label{Fig.Bperpnoise}
\end{figure}

More general expressions are discussed in Ref.~\cite{Khoo2021}. With our complete expressions for the conductivity from Eqs.~\eqref{Eq.condsppll}--\eqref{Eq.condchperp}, we have plotted the above autocorrelation function at low frequencies in Fig.~\ref{Fig.Bperpnoise} as a function of the NV center distance  $z$. We see that when the distance of the NV center is smaller than the spinon mean free path ($ p_{\rm F}^{-1} \ll z \ll l_{\rm mfp}$) the magnetic noise approaches the following universal value at low frequencies:
\begin{align}\label{Imchi}
{\rm Im} \chi _{B_z B_z} 
&\simeq \left( \frac{\chi_c}{\chi_s+\chi_{a}+\chi_c} \right)^2 \frac{e^2 \mu _0 ^2}{16 \pi h} \frac{\omega}{z}\frac{(2S+1)}{2\pi}\mathcal{P}_{\rm FS},
\end{align}
\noindent with subleading corrections to the above starting at $\mathcal{O}(\omega^3)$. The above formula applies to Fermi surfaces of arbitrary shapes and here $\mathcal{P}_{\rm FS}$ is the length of the perimeter of the spinon Fermi surface, which in the case of the circle is $\mathcal{P}_{\rm FS}=2\pi p_{\rm F}$. The above expression contains the dimensionless prefactor, $\chi_c^2/(\chi_s+\chi_{a}+\chi_c)^2$, which was missing in Ref.~\cite{Khoo2021}. From the above we have estimated the $T_1$ times in Table~\ref{Table.spinon} using the following formula~\cite{Casola2018,Langsjoen2012}:

\begin{align}\label{T1Eq}
\frac{1}{T_1}=\frac{\mu_B^2}{2\hbar} \coth \Bigl(\frac{\beta \hbar \omega }{2}\Bigr) {\rm Im} \chi _{B_z B_z}  (z, \omega),
\end{align}

\noindent where here we have simply set the magnetic moment of the NV to be the Bohr magneton for purposes of order-of-magnitude estimates. For Table~\ref{Table.spinon} we have taken $z=1$nm, $\omega=3$GHz and $T=10$K as parameters for the NV center~\cite{Casola2018}. The search for potential spin liquids states in 1T-TaS$_2$ and 1T-TaSe$_2$~\cite{law20171t, he2018spinon,ruan2021evidence,chen2021competition} might be simpler in monolayers compared to bulk samples, as there are two possible types of surfaces for the latter: unpaired single layer or paired double layer \cite{butler2020mottness}. Unfortunately so far the specific heat has only been measured in bulk samples, which  complicates the extraction of the spinon Fermi energy from bulk measurements, such as the specific heat~\cite{PhysRevB.96.195131}. This is why we have not indicated a value for $\epsilon_{\rm F}$ and $\Delta \omega_s$ in Table~\ref{Table.spinon}. Nevertheless, the estimate of the $T_1$ time can be performed without detailed knowledge of the spinon Fermi energy scale, thanks to its simple dependence on the length of the perimeter of the Fermi surface (which can be estimated from the spinon density).  This is why we have been able to provide an estimate for the expected $T_1$ time of 1T-TaS$_2$/Se$_2$ in Table~\ref{Table.spinon}.

We also would like to mention in passing that the discussion of this work does not necessarily apply to the ``pseudo-scalar" spinon Fermi surface introduced in Ref.~\onlinecite{PhysRevB.104.195149} to explain the oscillations of thermal conductivity in $\alpha$-RuCl$_3$~\cite{czajka2021oscillations,bruin2021robustness}.

\section{Summary and outlook}\label{Sec.summary}

We have shown that in the regime of wave-vectors $l^{-1}_{\rm mfp}\ll q \ll p_{\rm F}$, the dissipative part of the transverse electrical conductivity of a spinon Fermi surface state has the same form as that of a metal, but it is multiplied by the overall prefactor $\chi_c^2/(\chi_s+\chi_{a}+\chi_c)^2$, where $\chi_{s/c}$ are the spinon/chargon orbital diamagnetic susceptibilities and $\chi_{a}$ is the diamagnetic susceptibility of the emergent U(1) gauge field, correcting the results of Ref.~\cite{Khoo2021}. Interestingly this prefactor also controls the effective magnetic field that the spinons experience in response to a physical magnetic field and consequently the period of their quantum oscillations~\cite{motrunich2006orbital,chowdhury2018mixed,sodemann2018quantum}. 

The $1/T_1$ decay rate of an NV center placed near the spin liquid is therefore determined by the product of this prefactor and the length of the perimeter of the spinon Fermi surface, according to Eqs.~\eqref{Imchi}--\eqref{T1Eq}. We have estimated $T_1$ times for several spinon Fermi surface candidate materials and found that it ranges from a few tens to a few hundred milliseconds, as summarized in Table~\ref{Table.spinon}. 
The remaining challenge therefore, is to attain the experimental conditions required to measure these relatively long $T_1$ times. Nonetheless, NV centers are ideally suited for measuring the quasi-static regime of the transverse conductivity as they probe frequencies of the order of GHz, which are very small compared to the typical energy scales of these systems. 

Acknowledgements.   P.L. acknowledges the support by DOE office of Basic Sciences Grant No. DE-FG0203ER46076.

\bibliography{bibfile}

\clearpage

\appendix

\onecolumngrid

\section{Derivation of spinon Fermi surface state quasi-static transverse conductivity}\label{Sec.SFSScond}

The transverse conductivity in the quantum (collisionless) regime $q \gg q_*$ and at low frequencies $\omega\ll v_Fq$ can be approximated by 
\begin{align}
    \sigma_{\perp} (q,\omega)& \approx \biggl[\Bigl( \frac{p_{\rm F}}{2\pi q}+ (\chi_s+\chi_{a}) \frac{q^2}{i \omega}\Bigr)^{-1}+ \Bigl(\chi_c \frac{q^2}{i \omega}\Bigr)^{-1}\biggr]^{-1}\\
&\approx
    \frac{ \Bigl( \frac{p_{\rm F}}{2\pi q}+ (\chi_s+\chi_{a}) \frac{q^2}{i \omega}\Bigr)\Bigl(\chi_c \frac{q^2}{i \omega}\Bigr)}{
  \frac{p_{\rm F}}{2\pi q}+ (\chi_c+\chi_s+\chi_{a})\frac{q^2}{i \omega}}.
\end{align}
where we have set $g_S$, $\hbar$ and $e$ to one. The real part is a Lorentzian centered around zero frequency
\begin{align}
{\rm Re}\,    \sigma_{\perp} (q,\omega) \approx  \frac{p_{\rm F}}{2\pi q}
    \frac{ \chi_c^2}{
\Bigl(  \frac{p_{\rm F}\omega}{2\pi q^3}\Bigr)^2+ (\chi_c+\chi_s+\chi_{a})^2}.
\end{align}
At zero frequency this yields the result
\begin{align}
 {\rm Re}\,    \sigma_{\perp} (q,0) =
    \frac{ \chi_c^2}{  (\chi_c+\chi_s+\chi_{a})^2}\frac{p_{\rm F}}{2\pi q}.
\end{align}
The half width of the peak $ \Delta \omega_s$, for which $   \sigma_{\perp} (q,\Delta \omega_s)=  \sigma_{\perp} (q,0)/2$, is given by 
\begin{align}
 \Delta \omega_s=\frac{2\pi (\chi_c+\chi_s+\chi_{a})q^3}{p_{\rm F}}
\end{align}

\section{Derivation of the optical pseudo-gap}\label{Sec.Mottgap}

The conductivity in response to spatially uniform electric fields, obtained in the limit $\bf{q}\rightarrow 0$ of Eq.~\eqref{Eq.condsppll} or Eq.~\eqref{Eq.condspperp} of the main text, which governs optical and transport properties, is given by
\begin{align}
& \sigma (\omega) = \sigma _\| (0, \omega) = \sigma _\perp (0, \omega) = i \frac{n}{m} \frac{\omega (\omega^2 - \bar{\omega}_s^2 - i \omega \Gamma _1)}{(\omega^2 - \bar{\omega} ^2 - i \omega \Gamma _1)}, \label{Eq.condq0}\\
& \bar{\omega} = \sqrt{\frac{n }{m (\epsilon_c+\epsilon_s+\epsilon_{a})}}, \ \bar{\omega}_s = \sqrt{\frac{n }{m (\epsilon_s+\epsilon_{a})}}. \label{Eq.omegabar}
\end{align}
As shown in Fig.~\ref{Fig.condspinon}(b), this conductivity features a peak at $\omega = \bar{\omega}$. This peak can be viewed as the optical pseudo-gap or Mott optical lobe of these correlated states.

\section{Diamagnetic correction to Landau Fermi liquid theory}

Landau Fermi liquid theory is unable to properly capture the diamagnetic response of a Fermi liquid (see discussion around Eq.~[4.194] and Eq~[5.86] of Ref.~\onlinecite{Pines} as well as the absence of the diamagnetic response term in Eq~[8.75] of Ref.~\onlinecite{GiuVig}). As discussed in Ref.~\onlinecite{Pines}, one supplements the diamagnetic correction, obtained from a microscopic model (e.g. RPA), to the transverse current-current response tensor $\chi _\perp ^{\rm LFL} (q,\omega)$ obtained from Landau Fermi liquid theory.

More precisely, the diamagnetic term for a non-interacting electrons with $g_S$ spin components in an isotropic 2D system is given by~\cite{Pines,GiuVig}
\begin{equation}
\chi ^{\rm DM} _\perp (q, \omega \rightarrow 0) = g_S \frac{2}{3}\frac{n}{m} \left(\frac{q}{2p_{\rm F}}\right)^2 = g_S \frac{1}{6}\frac{n}{m} \frac{q^2}{p_{\rm F}^2},
\end{equation}
so that the full response function can be approximated as
%
\begin{equation}
\chi _\perp ^{\rm full} (q,\omega) \approx \chi _\perp ^{\rm LFL} (q,\omega)+ \chi ^{\rm DM} _\perp (q, \omega \rightarrow 0).
\end{equation}
Equivalently, through the relation between conductivity and response function, $\sigma (q, \omega) = e^2\chi (q, \omega)/i \omega$, one obtains the full transverse conductivity,
\begin{eqnarray}
\sigma _{\perp} (q, \omega) &\approx & \sigma _{\perp} ^{\rm LFL}(q, \omega) + g_S \frac{e^2}{i\omega}\frac{n}{6m} \frac{q^2}{p_{\rm F}^2}, \\
\sigma _{\perp} ^{\rm LFL} (q, \omega) &=& i\frac{n }{m} \frac{2}{ F_1 \omega _-  - \omega _+  - 2i \Gamma _2 }.
\end{eqnarray}
However, since the diamagnetic term is purely imaginary, it does not contribute to the dissipative transverse conductivity in metals. Consequently, the plots in Fig.~2(a) and Fig.~2(d) of Ref.~\onlinecite{Khoo2021}, and most importantly its main result on the universal transverse conductivity of metals, remain unchanged.

%
%
%

\end{document}